\documentclass[11pt,a4paper]{article}
\pdfoutput=1

\usepackage[colorlinks=true, linkcolor=black!50!blue, urlcolor=blue, citecolor=blue, anchorcolor=blue]{hyperref}
\usepackage[font=small,labelfont=bf,margin=0mm,labelsep=period,tableposition=top]{caption}
\usepackage[a4paper,top=2.5cm,bottom=3cm,left=2.3cm,right=2.3cm,bindingoffset=0mm]{geometry}

\usepackage{placeins,cite}
\usepackage{graphicx}
\usepackage{float}
\usepackage{afterpage}
\usepackage{epsfig}
\usepackage{amssymb}
\usepackage{amsmath}
\usepackage{bm}
\usepackage{multirow}
\usepackage{url}
\usepackage{xcolor}
\usepackage{ulem}
\usepackage{booktabs}
\usepackage{textcomp}
\usepackage[shortlabels]{enumitem}
\usepackage{tabularx}
\usepackage{subcaption}
\usepackage[capitalise]{cleveref}
\usepackage{bm}

\begin{document}

\begin{center}
  {\Large \bf Proton--Proton to Antinucleon Cross Sections 
    \\[0.25cm]
   for Cosmic Ray Applications
     }\\
  \vspace{1.1cm}
  {\small
    M. Boglione$^{1,2}$,
    M. Di Mauro$^{2}$, 
    F. Donato$^{1,2}$,
    E.R. Nocera$^{1,2}$,
    J. Rittenhouse West$^{2}$, 
    and A. Signori$^{1,2}$\\
  }
  
\vspace{0.7cm}

{\it \small
  ~$^1$Dipartimento di Fisica, Universit\`a degli Studi di Torino, Via Pietro
  Giuria 1, 10125 Torino, Italy\\
  ~$^2$INFN, Sezione di Torino, Via Pietro Giuria 1, 10125 Torino, Italy\\
  [0.1cm]
 }

\vspace{1.0cm}

\end{center}

\begin{abstract}
We present predictions of inclusive antiproton and antineutron production
cross sections in proton--proton collisions relevant to primary and secondary
antiproton production in cosmic ray interactions with interstellar matter.
Our predictions are based on collinear factorisation in Quantum Chromodynamics
and are accurate to next-to-leading order in the perturbative expansion of the
strong coupling. We assess the relevance of cross sections measured at collider 
experiments, such as NA49 at the CERN SPS and ALICE at the LHC, to the kinetic 
energy ranges accessed by cosmic ray detectors.  We characterise the
associated uncertainties due to the input parton distribution and fragmentation
functions, and to missing higher orders. We critically examine the $\sim$30\% 
excess of antineutron over antiproton production in proton--proton collisions 
preliminarily reported by the NA49 experiment by combining our predictions with
a data-driven  model. Our results do not support the NA49 finding, and point to
a mild excess  of a few percent. We finally show that the NA49 result could
only be reconciled with our framework by invoking sizeable differences between
antiproton and antineutron production in the poorly constrained region of small
transverse momenta of the produced hadron.
\end{abstract}
 
\section{Introduction}
\label{sec:intro}

Charged particles arriving at Earth from space are known as cosmic rays (CRs),
and those with energies above roughly 100 MeV are generally believed to
originate in and propagate through our Galaxy~\cite{Strong:2007nh,
  Gabici:2019jvz}. These are predominantly composed of protons, with minor
fractions of heavy nuclei, electrons, and antiparticles, such as antiprotons
and positrons. CR antiprotons are investigated as a potential indirect
signature of Galactic dark matter (DM) \cite{Donato:2003xg}, the existence
and abundance of which is supported by a variety of astrophysical and
cosmological evidence~\cite{Bertone:2010zza,Bertone:2016nfn,Cirelli:2024ssz}
indicating that gravitational phenomena cannot be described by baryonic matter
alone. In a broad class of scenarios, DM consists of weakly
interacting particles whose annihilation in the Galactic halo can produce
Standard Model (SM) particles. Indirect DM detection therefore focuses
on searching for such annihilation products as subdominant contributions to
astrophysical fluxes, particularly in cosmic antimatter channels such as
positrons, antiprotons, and antinuclei \cite{Salati:2010rc,Cirelli:2024ssz}.

The study of Galactic CRs has entered a precision era, driven by the
satellite-borne PAMELA experiment~\cite{PAMELA:2017bna} and, more recently, by
the AMS-02 detector operating on the International Space
Station~\cite{AMS:2021nhj}. In particular, AMS-02 has measured the CR
antiproton flux with an unprecedented, percent-level, precision for kinetic
energies from about 1 to 500~GeV~\cite{AMS:2015tnn,AMS:2016oqu,AMS:2021nhj,
  AMS:2025npj}. The observed flux is broadly consistent with a secondary
origin, namely with antiprotons produced by interactions of primary CRs
(essentially protons and helium nuclei) with the interstellar medium (ISM),
which is composed in turn mainly of hydrogen and helium
atoms~\cite{Donato:2001ms,Donato:2008jk,Boudaud:2019efq,DiMauro:2023jgg}.
Despite this remarkable precision, the current AMS-02 antiproton measurements
are not able to exclude the presence of a subleading DM contribution, due to
both experimental and theoretical uncertainties. In particular, probing DM with
CR antiprotons requires theoretical uncertainties in both the background and
the signal to be comparable to, or smaller than, the experimental uncertainties
of AMS-02~\cite{Donato:2017ywo,Maurin:2025gsz}.

Among the various sources of theoretical uncertainty, CR propagation in the
Galaxy plays a major role, affecting particles produced in the Galactic disk
and those originating in the extended halo in a different way, as expected for
DM annihilation~\cite{Donato:2003xg,Evoli:2011id,Genolini:2021doh,
  Calore:2022stf}. In this context, a key ingredient is the prediction of the
secondary source spectrum, which depends sensitively on the hadronic production
cross sections of CR protons and nuclei interacting with the
ISM~\cite{diMauro:2014zea,Winkler:2017xor,Donato:2017ywo,Korsmeier:2018gcy,
  Maurin:2025gsz} (see, {\it e.g.}, Eq.~(1) of~\cite{Korsmeier:2018gcy}).
Among these processes, proton--proton (\(pp\)) collisions provide the dominant
production channel and serve as the reference process to model reactions
involving heavier nuclei in the antiproton source term. A particularly
relevant theoretical uncertainty then arises from antineutron production,
\(pp \to \bar{n}X\), followed by the decay \(\bar{n}\to \bar{p}X'\), which
occurs on negligible timescales in the Galactic environment. In standard
calculations of the CR antiproton flux, the $pp \to \bar{n}X$ cross section is
typically assumed to be equal to that for prompt antiproton production,
\(pp \to \bar{p}X\)~\cite{Donato:2001ms}, and the antiproton source term is
conventionally rescaled by a factor of two to account for the contribution of
antineutron decays. However, preliminary results from the NA49 experiment
indicate that the antineutron yield may exceed the antiproton one by
approximately 30\%~\cite{Fischer:2003xh}. If confirmed, such an
asymmetry would have important implications for the interpretation of AMS-02
antiproton data. In particular, an uncertainty of this size in the antineutron
contribution propagates into an uncertainty of order 10\% in the total
secondary antiproton source term, to be compared with the percent-level
experimental precision achieved by AMS-02 over part of the measured energy
range. 

A common strategy to model the antiproton production cross sections in \( pp \)
collisions is to parametrise the Lorentz-invariant cross section and fit the
free parameters to experimental data, see {\it e.g.}~\cite{diMauro:2014zea,
  Winkler:2017xor,Donato:2017ywo, Korsmeier:2018gcy}. This approach provides an
accurate description in the kinematic phase space covered by measurements.
Its main limitation, however, is that it remains model-dependent and is
therefore not universal. Moreover, the extrapolation to poorly measured
kinematic regions or to production channels that have not been directly
measured may not be robust. Therefore this method does not provide a unified
theoretical framework for inclusive hadron production cross sections across
different species and kinematic regimes.

An alternative approach relies on Monte Carlo event generators, such as
{\sc Pythia}~\cite{Skands:2014pea}, which  provide a fully exclusive
description of the final state. However, in their standard parameter sets
(tunes) they do not accurately reproduce the antiproton yield at fixed-target
and intermediate energies relevant for CR studies, thus requiring dedicated
retuning~\cite{diMauro:2026oto}. Moreover, {\sc Pythia} predicts equal
production rates for antiprotons and antineutrons in \( pp \) collisions,
whereas the aforementioned NA49 measurements~\cite{Fischer:2003xh} suggest that
the two yields may differ. Finally, the modeling of collisions involving nuclei
heavier than hydrogen is subject to larger theoretical and phenomenological
uncertainties, limiting the reliability of these predictions for CR
applications. Very recently, an alternative class of hadronic interaction
models based on Reggeon field theory has been
developed~\cite{Ostapchenko:2026uds}, specifically tailored to CR applications.
These models are implemented in Monte Carlo generators that provide a
computationally efficient and physically transparent description of hadronic
interactions, while preserving basic theoretical constraints such as unitarity
and conservation laws and allowing for phenomenological retuning. Such
approaches offer a complementary framework to the methods discussed above.
However, similarly to other Monte Carlo generators, they do not provide a fully
controlled description of possible differences between antiproton and
antineutron production, which are sensitive to the underlying flavour structure
of the hadronisation process, among other effects.

Despite these developments, achieving a consistent and theoretically controlled
description of antiproton and antineutron production, including their possible
differences, over the full kinematic range relevant to CR applications
remains challenging. We therefore investigate a third
approach: we compute the Lorentz-invariant cross section for inclusive hadron
production in \( pp \) collisions within perturbative Quantum Chromodynamics
(QCD) and collinear factorisation. In its domain of
applicability, this approach is universal, and it provides a theoretically
controlled description of the underlying strong interaction dynamics of the
process. It enables us to investigate potential differences
between antiproton and antineutron production
in a framework where the flavour dependence of the
hadronisation process is explicitly encoded in parton distribution (PDFs) and
fragmentation functions (FFs). This framework also allows for a systematic
assessment of the theoretical uncertainties entering the calculation, including
those associated with PDFs and FFs, as well as with missing higher-order
corrections in the perturbative expansion of the strong coupling.
Fixed-order predictions are, however, not reliable over the full range of the
transverse momentum $p_T$ of the observed hadron, since the factorised
expression for the cross section receives ${\mathcal O}(m/p_T)$ corrections,
where $m$ is the proton mass. Thus, for $p_T \lesssim 1$~GeV resummation and
non-perturbative effects become important and the fixed-order calculation
progressively loses its predictive power. Moreover, a factorised expression 
for the cross section at small transverse momentum $p_T \lesssim 1$~GeV 
in terms of transverse-momentum-dependent partonic distributions cannot be 
formally achieved for inclusive production of hadrons from hadronic 
collisions~\cite{Collins:2007nk,Rogers:2010dm}.
For these reasons, the fixed-order result valid at $p_T > 1$~GeV 
must be matched to a phenomenological parametrisation constrained by
existing data in the low-$p_T$ region, where the measurements are typically
most precise.

This hybrid strategy combines the strengths of two approaches. At small $p_T$,
it preserves the direct experimental information encoded in traditional
data-driven parametrisations. At larger $p_T$, it replaces unconstrained
parametric extrapolations with a prediction grounded in perturbative QCD.
This is advantageous even though the CR antiproton spectrum is obtained by
integrating the Lorentz-invariant cross section over the entire production
solid angle and receives its dominant contribution from the
low-$p_T$ region~\cite{Donato:2017ywo,Korsmeier:2018gcy}. The uncertainty
associated with the less constrained high-\(p_T\) tail, where the cross section
decreases steeply with $p_T$, is indeed still relevant after integration over
the phase space. It is therefore crucial to estimate it correctly.
Our framework enables a critical assessment of whether the sizeable antineutron
excess suggested by NA49 is compatible with a QCD-based description,
supplemented with data in the region where perturbation theory is not
applicable.

The paper is organised as follows. In Sect.~\ref{sec:xsec}, we define the
physical cross section relevant for modeling primary and secondary antiproton
production in \( pp \) collisions, discuss how it can be computed in
perturbative QCD, and show how measurements performed by collider experiments,
such as NA49 and ALICE, relate to measurements performed by cosmic ray
detectors. In Sect.~\ref{sec:apanzsecs}, we revisit the claim made by the NA49
experiment of an excess of antineutron over antiproton inclusive production
yields in \( pp \) collisions. To this purpose, we lay out a prescription
to reconstruct and integrate the predicted Lorentz-invariant cross section over
the entire range of $p_T$, separately for the production of antiprotons and
antineutrons, and carefully assess all the uncertainties involved. Finally, in
Sect.~\ref{sec:conclusions}, we summarise our main results and discuss their
implications for CR antiproton modelling. Our paper is completed by two
appendices. In Appendix~\ref{app:ALICE}, we present a validation of our
computations against ALICE measurements, which are known to be well described
by the perturbative QCD framework. In Appendix~\ref{app:pions}, we extend our
computations to the case of positively and negatively charged inclusive pion
production in \( pp \) collisions, and compare them to the NA49
measurements~\cite{NA49:2009brx}, where a slight excess of the cross section
for the production of positively over negatively
charged pions is reported, in particular at low $p_T$.

\section{Cross section framework and kinematic coverage}
\label{sec:xsec}

In this section, we first discuss the computation of the cross section for the 
scattering processes relevant to this work, including details on their QCD 
factorisation and numerical implementation. We then show how the corresponding
measurements performed by the NA49~\cite{NA49:2009brx} and 
ALICE~\cite{ALICE:2014juv,ALICE:2019hno,ALICE:2020jsh} experiments are
sensitive to the nonperturbative input, and how they relate to measurements
performed  by CR detectors.

\subsection{Perturbative computation}
\label{subsec:pert_comp}

The high-energy scattering processes of interest to this paper are the
inclusive production of an antiproton in \(pp\) collisions, and the
secondary production of an antiproton from the decay of an antineutron, in turn
produced inclusively in \(pp\) collisions. These processes read,
respectively, as
\begin{equation}
  pp\to\bar p(E,{\bm p}) X
  \qquad
  \qquad
  pp\to\bar n(E,{\bm p}) X \to \bar p X^\prime\,,
  \label{eq:process}
\end{equation}
where we have explicitly denoted the energy $E$ and the three-momentum $\bm p$ 
of the produced antiproton (or intermediate antineutron). Because an
antineutron decays into an antiproton, a positron, and an antineutrino with a
decay width close to unity~\cite{ParticleDataGroup:2024cfk} and since the 
antiproton and antineutron momenta are  equivalent  to the per-mille level, 
the antiproton cross section for the second process in Eq.~\eqref{eq:process} 
is effectively equivalent to that for the inclusive production of an
antineutron in \(pp\) collisions.

In QCD, the corresponding Lorentz-invariant cross sections factorise in terms
of a perturbative hard coefficient, $\hat\sigma_{ab}^c(\mu_R^2,\mu_F^2,\mu_f^2)$,
initial-state PDFs, $f_a(x_1,\mu_F^2)$ and
$f_b(x_2,\mu_F^2)$, and a final-state FF,
$D_c^h(z,\mu_f^2)$:
\begin{equation}
  E\frac{d^3\sigma}{dp^3}
  = \sum_{a,b,c}
  \hat\sigma_{ab}^c(\mu_R^2,\mu_F^2,\mu_f^2)
  \otimes f_a(x_1,\mu_F^2)
  \otimes f_b(x_2,\mu_F^2)
  \otimes D_c^h(z,\mu_f^2)\,.
  \label{eq:Lorentz_xsec}
\end{equation}
In Eq.\eqref{eq:Lorentz_xsec}, the indexes $a$, $b$, and $c$ denote all the
active partons at a given scale, $h$ denotes the species of the final-state
hadron (in the case of interest, an antiproton or an antineutron), $x_1$ and
$x_2$ are the fractions of the initial-state proton momenta
carried by the partons involved in the hard scattering, $z$ is the momentum
fraction of the fragmenting parton carried by the produced hadron, $\mu_F$
and $\mu_f$ are the PDF and FF factorisation scales, respectively, $\mu_R$ is
the renormalisation scale,  and $\otimes$ is the usual convolution product
\begin{equation}
  g(x)\otimes h(x)=\int_x^1\frac{dz}{z}g\left(\frac{x}{z} \right) h(z)\,.
  \label{eq:convolution}
\end{equation}
The perturbative expansion in the strong coupling $\alpha_s$ of the partonic
cross section $\hat\sigma_{ab}^c(\mu_R^2,\mu_F^2,\mu_f^2)$ has been known for a
long time up to next-to-leading order (NLO)~\cite{Aversa:1988vb,Aurenche:1999nz,
  deFlorian:2002az,Jager:2002xm} and has been computed very recently up to
next-to-next-to-leading order (NNLO)~\cite{Czakon:2025yti}. The fixed-order
computation is supposed to be adequate for values of the transverse momentum of
the produced hadron $p_T$ larger than about 1~GeV. Below this scale,
resummation and non-perturbative effects hinder the accuracy of
Eq.~\eqref{eq:Lorentz_xsec}.

The appropriate comparison between the theoretical predictions computed using
Eq.~\eqref{eq:Lorentz_xsec} and the experimental data depends on the details of
each measurement. The NA49 experiment~\cite{NA49:2009brx} measured the cross
section in Eq.~\eqref{eq:Lorentz_xsec} for inclusive antiproton production
in bins of $x_F$ and $p_T$; in the centre-of-mass frame, $x_F=2p_L/\sqrt{s}$,
with $p_L$ and $p_T$ the longitudinal and transverse components of the
antiproton three-momentum. The antiproton rapidity
\begin{equation}
  y=\frac{1}{2}\ln\left(\frac{E+p_L}{E-p_L}\right)\,
  \label{eq:rapidity}
\end{equation}
can be rewritten in terms of $x_F$ as
\begin{equation}
  y={\rm arcsinh}\left(\frac{\sqrt{s}}{m_T}\frac{x_F}{2}\right)\,
  \label{eq:rapidity_xF}
\end{equation}
by recalling that $E^2=p_L^2+m_T^2$, where $m_T=\sqrt{p_T^2+m^2}$ is the
transverse mass, and $m=m_{\bar p}=938.27$~MeV is the mass of the antiproton. It
follows that $x_F=0$ corresponds to $y=0$, and likewise that $x_F>0$ ($x_F<0$)
corresponds to forward (backward) rapidity. To match the NA49 measurements, it
is therefore sufficient to evaluate Eq.~\eqref{eq:Lorentz_xsec} for fixed
values of $p_T$ and $y$, computed from values of $x_F$ according to
Eq.~\eqref{eq:rapidity_xF}. Conversely, the ALICE
experiment~\cite{ALICE:2020jsh} measured the cross section in
Eq.~\eqref{eq:Lorentz_xsec} for the sum of inclusive proton and
antiproton production in bins of $p_T$, with rapidity integrated
over $|y|\leq 0.5$. To match the ALICE measurements, one must therefore
evaluate Eq.~\eqref{eq:Lorentz_xsec} for fixed values of $p_T$ and rapidity
integrated over the specified interval.

We compute predictions for the cross section in Eq.~\eqref{eq:Lorentz_xsec}
by means of the code developed in~\cite{deFlorian:2002az}, which we
have interfaced to {\sc PineAPPL}~\cite{Carrazza:2020gss}, see
also~\cite{Cruz-Martinez:2025ahf}. This allows us to precompute the
perturbative part of the cross section on a discretised and suitably optimised
interpolation grid. The numerical evaluation of the convolution integral in
Eq.~\eqref{eq:Lorentz_xsec} then becomes very fast, as it reduces to a
weighted sum of interpolating functions evaluated on the grid nodes.
Other public software packages exist in the literature, namely
{\sc INCNLO}~\cite{Aurenche:1998gv} and {\sc FMNLO}~\cite{Liu:2023fsq},
however they are not interfaced to {\sc PineAPPL}.
The accuracy of our computations is NLO in the strong coupling $\alpha_s$.
As mentioned, NNLO corrections have been recently computed, however these are
not delivered in a format that we can readily include in our computations.
We set $\mu_F^2=\mu_f^2=\mu_R^2=p_T^2$. The impact of missing higher-order
corrections can be studied by variation of the scales $\mu_F$,
$\mu_f$, and $\mu_R$ by a factor $0.5$ or $2.0$, for instance, by taking
the envelope of the 16-point scale variations, which exclude the pairwise
variation of two of the three scales by a factor $0.5$ or $2.0$ (see
{\it e.g.} Eq.~(2) in~\cite{dEnterria:2013sgr}). We will discuss the size of
these uncertainties for the NA49 measurement, in comparison to PDF and FF
uncertainties, in Sect.~\ref{subsec:antiproton_NA49}.

We use the {\sc NNPDF4.0} proton PDF set~\cite{NNPDF:2021njg} and the
{\sc NNFF1.0} antiproton FF set~\cite{Bertone:2017tyb}, specifically the
baseline NLO PDF and FF sets, corresponding to a value of the strong coupling
at the $Z$-mass pole $\alpha_s(M_Z)=0.118$. Other antiproton FF sets exist in
the literature, among which {\sc KKP}~\cite{Kniehl:2000fe},
{\sc AKK}~\cite{Albino:2005me}, {\sc HKNS}~\cite{Hirai:2007cx},
{\sc DSS}~\cite{deFlorian:2007ekg}, and {\sc NPC23}~\cite{Gao:2024dbv}.
However, all of them but {\sc NPC23} are not available in the
{\sc LHAPDF} format~\cite{Buckley:2014ana}, a fact that makes their usage for
the computation of Eq.~\eqref{eq:Lorentz_xsec} technically complicated, and
prevents an assessment of the uncertainties due to FFs.
On the other hand, the {\sc NPC23} FF set covers energy values of $Q=\mu_f$ only
larger than 4~GeV. This fact implies that extrapolation to lower values of
$Q=p_T\sim 1$~GeV, would be needed for a comparison with the NA49
experimental data. The {\sc NNFF1.0} set overcomes all these limitations:
this FF set is publicly available in the {\sc LHAPDF} format, and it covers a
kinematic range down to $Q=1$~GeV. We consistently choose the {\sc NNPDF4.0}
proton PDF set, that belongs to the same family of fits and is specifically
based on the same fitting methodology. We take the so-called perturbative charm
PDF set, in which there is no intrinsic component of the charm PDF, which also 
covers a kinematic range down to $Q=1$~GeV. Unless otherwise specified, in all
of our computations we keep the input PDF fixed to the central value, and we
vary the FF among all of the 100 Monte Carlo FF replicas in the NNFF1.0 
set. This amounts to neglecting the PDF uncertainty on the cross section,
the size of which we will compare to FF and scale uncertainties in
Sect.~\ref{subsec:antiproton_NA49}. We finally construct the antineutron FF,
required to compute the second process in Eq.~\eqref{eq:process}, by applying
exact isospin symmetry to the antiproton FFs.

We have validated our theoretical predictions against the ALICE antiproton
measurements~\cite{ALICE:2014juv,ALICE:2019hno,ALICE:2020jsh} included in the
{\sc NPC23} analysis, using consistently their PDFs and FFs. In this case,
our predictions actually become postdictions, that must match the data as
in~\cite{Gao:2024dbv}. This validation is illustrated
in Appendix~\ref{app:ALICE}. We have also verified that the {\sc NNFF1.0}
set, despite not including the aforementioned data, correctly predicts
the measured large-$p_T$ behaviour of the cross section. This sanity check,
also presented in Appendix~\ref{app:ALICE}, corroborates the reliability of
our chosen non-perturbative input to correctly describe the cross section
over a broad $p_T$ range.

\subsection{Quark and gluon fragmentation contributions}
\label{subsec:sensitivity}

\begin{figure}[!t]
  \centering
  \includegraphics[width=0.49\textwidth]{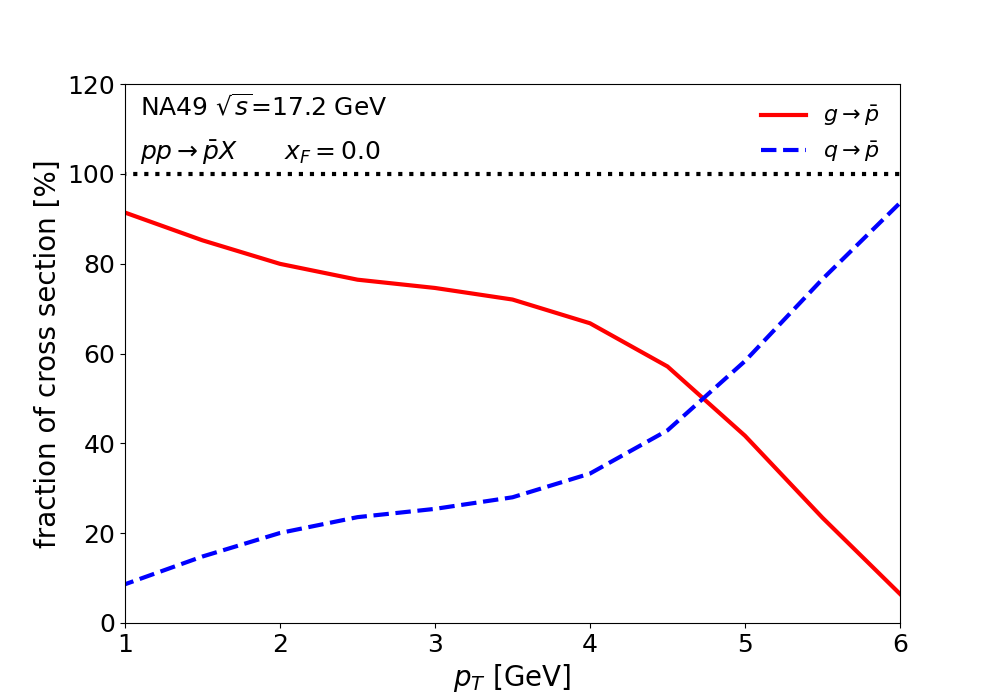}
  \includegraphics[width=0.49\textwidth]{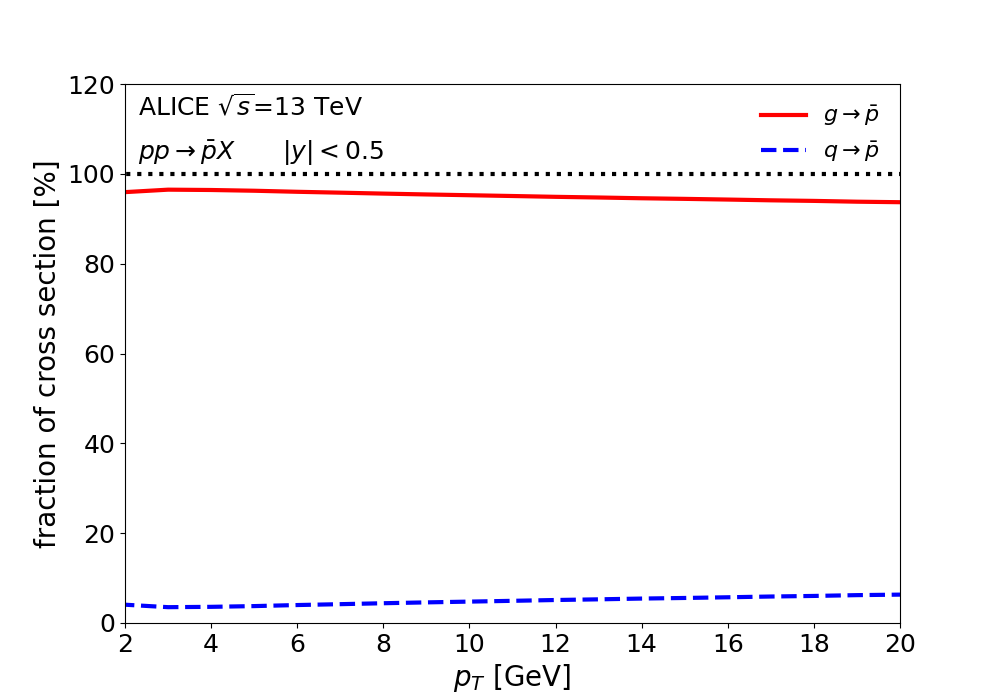}\\
  \caption{The percentage fraction of the cross section,
    Eq.~\eqref{eq:Lorentz_xsec}, defined as the ratio between the 
    contribution to the cross section of a partonic species and the total 
    cross section. The gluon fragmentation channel ($g\to \bar p$) is 
    indicated with a dashed blue line. The contribution to the cross section 
    from the fragmentation of
    all active quarks ($q\to\bar p$) is indicated with a solid red line. 
    Results are displayed as a function of $p_T$ for two 
    representative experiments that cover different kinematic regions: 
    NA49~\cite{NA49:2009brx} ($\sqrt{s}=17.2$~GeV, left panel), and
    ALICE~\cite{ALICE:2020jsh} ($\sqrt{s}=13$~TeV, right panel). 
    Results are obtained at NLO accuracy using the central replica of
    the {\sc NNPDF4.0}~\cite{NNPDF:2021njg}
    and {\sc NNFF1.0}~\cite{Bertone:2017tyb} NLO PDF and FF sets, see text for 
    details.} 
    \label{fig:channels}
\end{figure}

The Lorentz-invariant cross section, Eq.~\eqref{eq:Lorentz_xsec}, receives
contributions from quark- and gluon-initiated fragmentation into the
final-state antiproton. In Fig.~\ref{fig:channels}, we display the relative
size of each of the two contributions, for two representative centre-of-mass
energies $\sqrt{s}$: 17.2~GeV, corresponding to the NA49
experiment~\cite{NA49:2009brx}, and 13~TeV, corresponding to the ALICE
experiment~\cite{ALICE:2020jsh}. The fraction of cross section is defined as
the ratio of the contribution to the cross section from either quark
($q\to\bar p$) or gluon ($g\to \bar p$) fragmentation channels into the final
antiproton to the total cross section. Our computations are performed as
explained above, with the exception that we take only the central value of the
NNFF1.0 set. We observe that gluon fragmentation dominates across the range
of transverse momenta $p_T$ covered by the corresponding experiments. This
fact suggests, as we will further see in Sect.~\ref{sec:apanzsecs}, that the
cross sections for the inclusive production of an antiproton or an antineutron
are expected to be very similar. By construction, the two only differ for quark
FFs, which however contribute to less than 20\% to the total cross section in
the $p_T$ range relevant to the NA49 measurements.

\subsection{Phase-space mapping and experimental coverage}
\label{subsec:phase-space}

The computation of the cross section in Eq.~\eqref{eq:Lorentz_xsec} equally
concerns measurements performed at colliders, such as NA49 and ALICE, or by
CR detectors. Actually, the former have been sometimes used to inform
the latter~\cite{Donato:2017ywo}, {\it e.g.}, in the case of the excess of 
antineutron over antiproton production reported by NA49~\cite{Fischer:2003xh}.
The reason is that collider measurements and CR events have an overlapping
kinematic coverage. Working with natural units, the kinetic energy in the
laboratory frame is related to the energy of the final-state antiproton
produced inclusively in the \(pp\) collision as 
\begin{equation}
T_{\rm lab}=E_{\rm lab}-m\,,
\label{eq:Tlab}
\end{equation}
where
$E_{\rm lab}=\gamma(E+\beta p_L)$ and $E_{\rm lab}=E$, 
\label{eq:Elab}
for fixed-target (such as NA49) and collider (such as ALICE)
experiments, respectively. The $\gamma$ and $\beta$ factors are the boost 
parameters of the centre-of-mass frame with respect to the laboratory frame: 
\begin{equation}
  \beta = v \, , 
  \qquad
  \gamma = \frac{1}{\sqrt{1-\beta^2}} = \frac{E_{\rm beam} + m_p}{\sqrt{s}} \, ,
  \label{eq:boost_pars}
\end{equation}
where $v$ is the relative velocity of the two frames, $E_{\rm beam}$ is the 
energy of the incident beam, and $m_p$ is the mass of the proton target. 
We therefore obtain
\begin{equation}
  T_{\rm lab}=\gamma\left(\sqrt{\frac{x_F^2}{4}s+m_T^2} + \beta\frac{x_F}{2}\sqrt{s} \right) - m
  \qquad
  T_{\rm lab}=\sqrt{\frac{x_F^2}{4}s+m_T^2}-m\,,
  \label{eq:bounds_on_T}
\end{equation}
respectively for fixed-target (NA49) and collider (ALICE) experiments. 
For fixed values of the kinetic energy, Eqs.~\eqref{eq:bounds_on_T} identify a 
corresponding region in $x_F$ and $p_T$ that can then be compared to the
kinematic region typically accessed by colliders. Note that CR production in
the Galaxy occurs in the laboratory frame, where the parent CR nucleus acts as
the beam and the ISM atom as the target.

\begin{figure}[!t]
  \centering
  \includegraphics[width=0.49\textwidth]{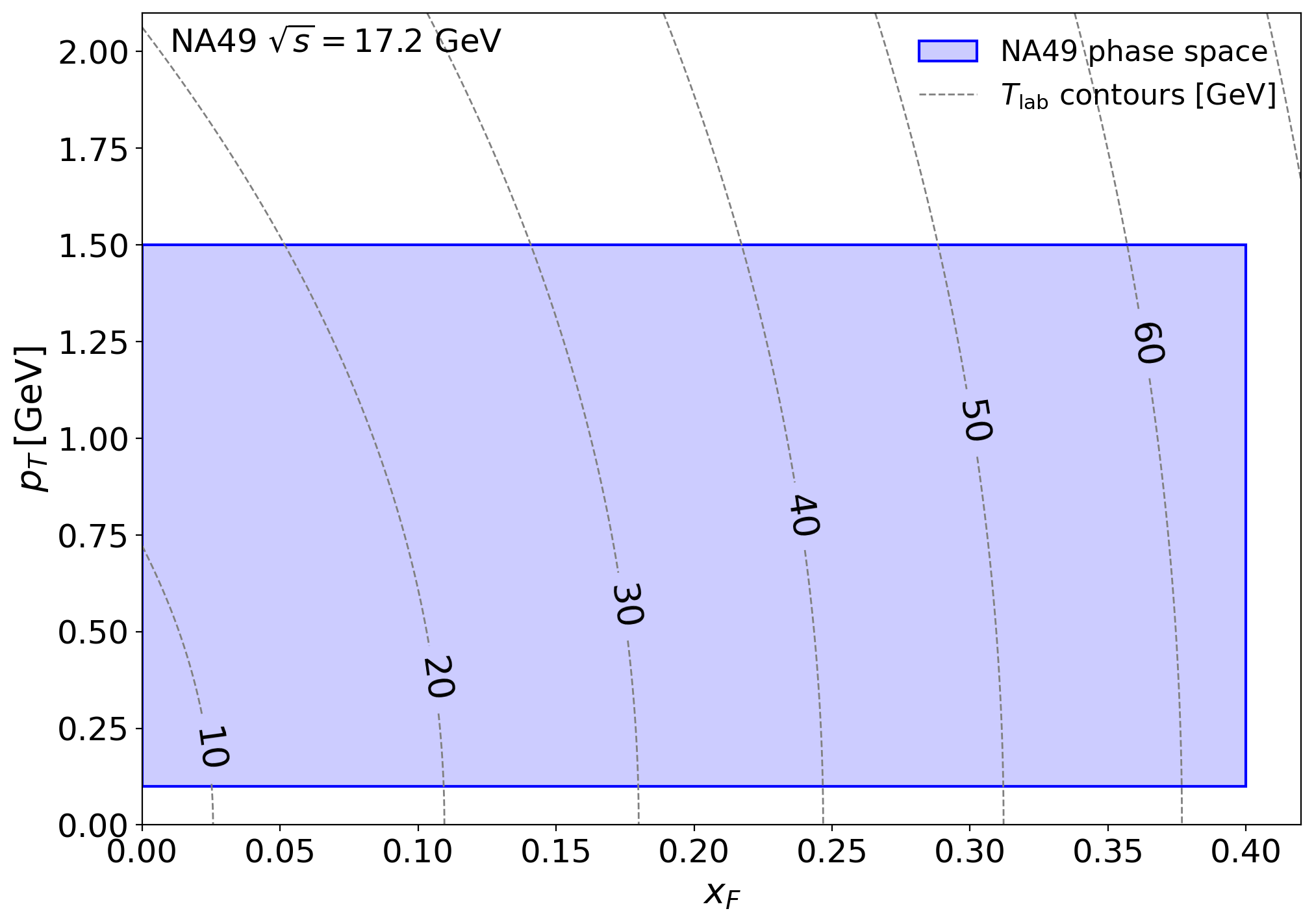}
  \includegraphics[width=0.49\textwidth]{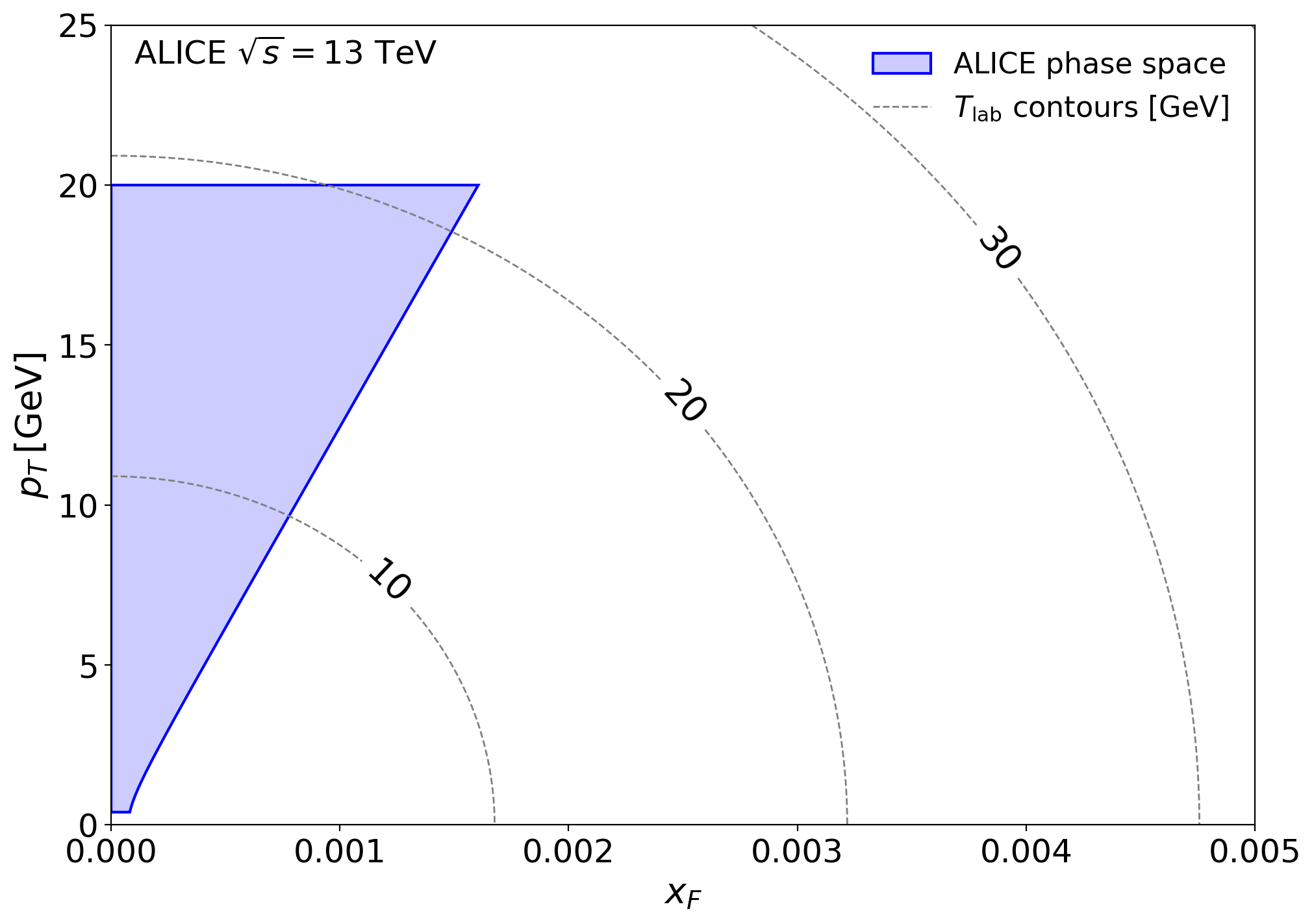}\\
  \caption{Fixed-target NA49 (left) and collider ALICE (right) phase space, in 
  the ($x_F,p_T$) plane, compared to contours of fixed kinetic energy 
  $T_{\rm lab}$ (Eqs.~\eqref{eq:bounds_on_T}). For ALICE, the range in $x_F$ is 
  obtained by inverting Eq.~\eqref{eq:rapidity_xF} and evaluating it for
  $y=\pm 0.5$.
  }
  \label{fig:bounds}
\end{figure}

In Fig.~\ref{fig:bounds} we display the phase-space region covered by the NA49
and ALICE experiments in the ($x_F,p_T$) plane. 
For ALICE, the corresponding range in $x_F$ is obtained by inverting 
Eq.~\eqref{eq:rapidity_xF} and evaluating it for $y = \pm 0.5$
(see~\cite{ALICE:2020jsh} and Appendix~\ref{app:ALICE}). 
We also show contours for Eqs.~\eqref{eq:bounds_on_T}. We remind that the
antiproton energies measured by AMS-02 range from approximately 1~GeV to
500~GeV. As a very rough rule, the corresponding parent CR nucleus carries an 
energy  $T_{\rm lab}$ which is on average  10-20 times
higher~\cite{Korsmeier:2018gcy}. This confirms that accelerator experiments
probe a laboratory energy range relevant for CR measurement in space. 

\section{The antineutron vs antiproton production excess}
\label{sec:apanzsecs}

In this section, we address the claim made by the NA49
experiment~\cite{Fischer:2003xh} of an excess of antineutron over antiproton
inclusive production yields in \(pp\) collisions. We first compute
perturbative QCD predictions for single-inclusive antiproton production, using
the framework discussed in Sect.~\ref{subsec:pert_comp}. We assess the size
of PDF, FF, and scale uncertainties, and compare them to the NA49
measurements~\cite{NA49:2009brx}. We then lay out a prescription to reconstruct
and integrate the Lorentz-invariant cross section over the entire range of
$p_T$, separately for the production of antiprotons and antineutrons. We
finally present results for the antineutron and antiproton integrated production
yields, including their ratio, we assess their uncertainties, and compare them
to the original NA49 result~\cite{Fischer:2003xh}. We comment on how our
theoretical predictions can be modified to reconcile with that result.

\subsection{Antiproton production at NA49}
\label{subsec:antiproton_NA49}

The NA49 experiment~\cite{NA49:2009brx} measured the Lorentz-invariant cross
section for the inclusive production of protons and antiprotons in
\(pp\) collisions. The fixed-target detector at the CERN SPS collected
4.8 million inelastic events with a proton beam momentum of 158 GeV,
corresponding to a centre-of-mass-energy of 17.2~GeV. The Lorentz-invariant
cross section was delivered in bins of the tranverse momentum of the detected
hadron, $p_T$, and of $x_F$, as defined in Sect.~\ref{subsec:pert_comp}. For
antiprotons, the covered intervals for these two kinematic variables were
$p_T\in[0.1, 1.5]$~GeV and $x_F\in[-0.05, 0.40]$.
On the other hand, the previous NA49 pilot run~\cite{Fischer:2003xh} measured a
small sample of 120,000 events of antiproton production in \(pn\)
collisions at the same centre-of-mass energy. The Lorentz-invariant cross
section was presented in bins of $x_F$, with $x_F\in[-0.05, 0.25]$, integrated
over $p_T$. This result, which showed a sizeable increase of antiproton
production in \(pn\) collisions in comparison to \(pp\)
collisions for $x_F=0$, has been used as a proxy for the inclusive production
of an antineutron in \(pp\) collisions by means of crossing symmetry.

We start to assess this state of affairs by computing theoretical predictions
for the inclusive production of an antiproton or an antineutron in
\(pp\) collisions, using the framework of Sect.~\ref{subsec:pert_comp}.
Given that the aforementioned production excess was observed for $x_F=0$, we
restrict ourselves to this value.

\begin{figure}[!t]
  \includegraphics[width=0.49\textwidth]{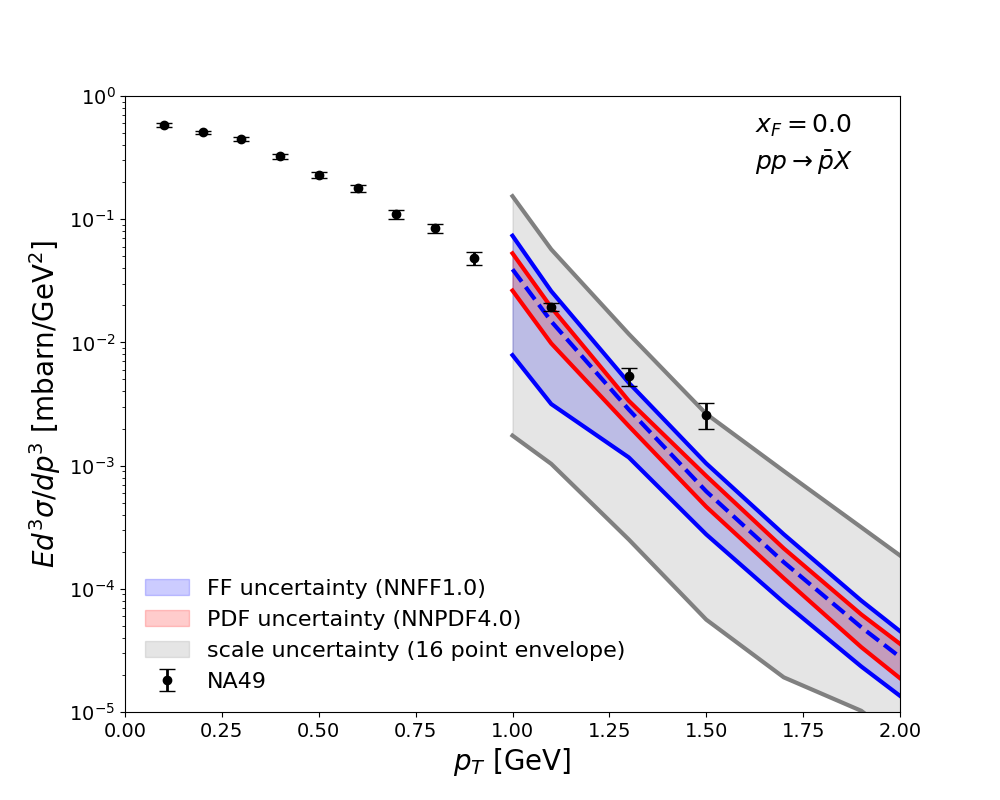}
  \includegraphics[width=0.49\textwidth]{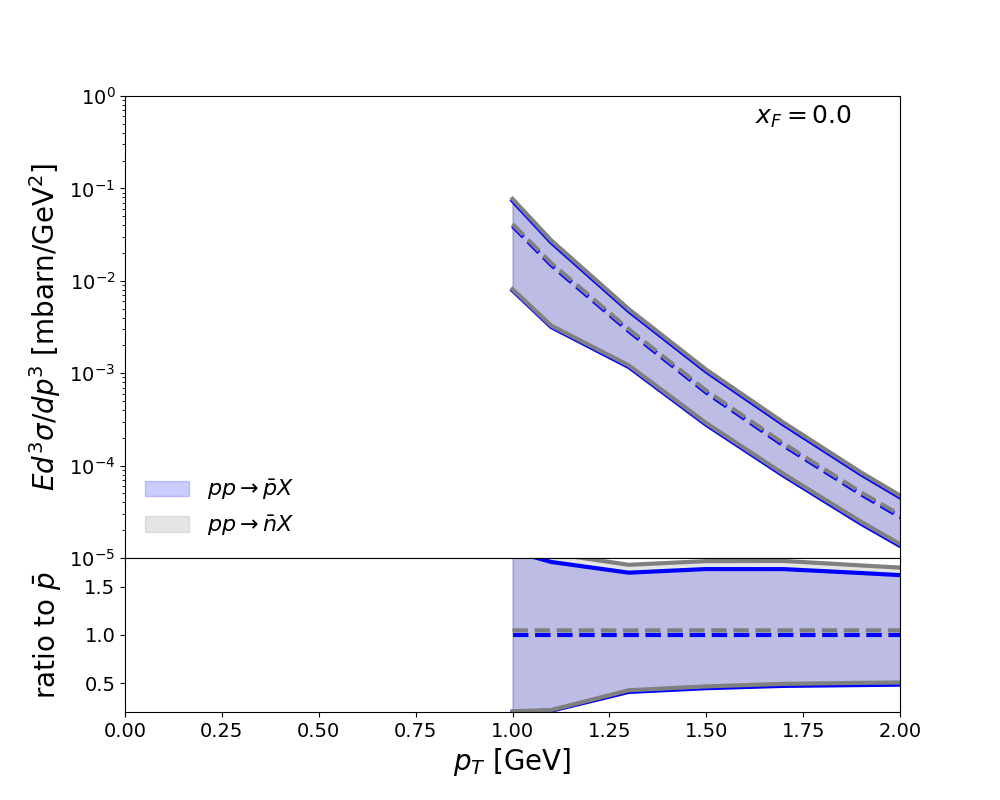}\\
  \caption{(Left) Comparison of our theoretical predictions, computed as
    explained in Sect.~\ref{subsec:pert_comp}, to the NA49 experimental
    measurements of~\cite{NA49:2009brx}.  We separately indicate the FF, PDF,
    and scale uncertainties on theoretical predictions. The FF (PDF)
    uncertainty corresponds to the 68\% confidence level computed over the FF
    (PDF) Monte Carlo ensemble of replicas provided by the nominal NNFF (NNPDF)
    parton sets, keeping the PDF (FF) fixed to its corresponding average.
    The scale uncertainty corresponds to the envelope of the 16-point scale
    variations. The experimental uncertainties of the NA49 measurement are the
    sum in quadrature of all statistical and systematic uncertainties. We do
    not display theoretical predictions for a value of $p_T$ below 1~GeV, as
    perturbative QCD is no longer applicable in that region.
    (Right) Comparison of theoretical predictions for the inclusive
    production of an antineutron and of an antiproton (the latter are the same
    as in the left plot), normalised to the latter (bottom panel). We display
    only the FF uncertainty, PDF and scale uncertainties remaining the same as
    for antiprotons. We do not display any experimental measurements, as no 
    measurements differential in $p_T$ are available.}
  \label{fig:antiprotons}
\end{figure}

We first focus on antiprotons. In the left plot of Fig.~\ref{fig:antiprotons},
we compare our theoretical predictions to the NA49 experimental measurements
of~\cite{NA49:2009brx}. We do not display theoretical predictions for
values of $p_T$ below 1~GeV, as perturbative QCD is no longer applicable in
that region. We separately indicate the FF, PDF, and scale uncertainties on
theoretical predictions. The FF (PDF) uncertainty corresponds to the 68\%
confidence level computed over the FF (PDF) Monte Carlo ensemble of replicas
provided by the nominal NNFF (NNPDF) parton sets, keeping the PDF (FF)
fixed to its corresponding central value. The scale uncertainty corresponds to
the envelope of the 16-point scale variations, obtained as explained in
Sect.~\ref{subsec:pert_comp}. The experimental uncertainties of the NA49
measurement are the sum in quadrature of all statistical and systematic
uncertainties.

We observe that our predictions, where applicable, describe the NA49
measurements relatively well; in particular, the large-$p_T$ data points are
always compatible with our theoretical predictions within their rather large
uncertainties. In this respect, we note the expected hierarchy of uncertainties:
the PDF uncertainty is the smallest, consistently with the fact that PDFs are
nowadays very well constrained by the data; then comes the FF uncertainty,
which is two to five times larger than the PDF uncertainty, as expected from
the fact that FFs are significantly less constrained than PDFs; and finally the
scale uncertainty is up to one order of magnitude larger than that, consistently
with the fact that perturbative corrections can be large at such small values of
$p_T$. We note that the quantity we are interested in is the ratio
for the inclusive production of antineutrons to antiprotons, integrated over
$p_T$. Because the structure of the cross section is the same in the two cases,
PDF and scale uncertainties behave exactly the same, and are 100\%
correlated. As we will see in Sect.~\ref{subsec:ratio_yields}, this fact
results in their almost exact cancellation.

We then look at antineutrons. In the right plot of Fig.~\ref{fig:antiprotons},
we compare theoretical predictions for the inclusive production of an
antineutron and of an antiproton (the latter are the same as in the left plot
in Fig.~\ref{fig:antiprotons}). For simplicity, we display only the FF
uncertainty, given that PDF and scale uncertainties remain the same as for
antiprotons. We do not display any experimental measurement, given that none
exist differential in $p_T$. As expected, the two sets of predictions differ
very little between each other. As observed in Sect.~\ref{subsec:sensitivity},
they both receive their leading contribution from gluon fragmentation, which is
by construction the same for an antiproton and for an antineutron. The observed
3-5\% difference follows from the different quark FFs, which however only
contribute to the 20\% of the cross section or less (see 
Fig.~\ref{fig:channels}). This fact suggests that
any difference between the cross section for the inclusive production of an
antiproton or an antineutron should come from the small-$p_T$ region, as we will
further emphasise in Sect.~\ref{subsec:ratio_yields}.

\subsection{Reconstruction of the small-$p_T$ behaviour}
\label{subsec:low_pT}

The NA49 experiment originally reported an excess of antineutron over
antiproton production~\cite{Fischer:2003xh} by measuring the
Lorentz-invariant cross section, Eq.~\eqref{eq:Lorentz_xsec}, integrated over
the entire range of $p_T$ for a fixed value $x_F=0$. In order to assess this 
claim we therefore compute
\begin{eqnarray}
  \frac{d\sigma}{dx_F}
  & = & 
  \int_0^\infty dp_T^2\frac{d^2\sigma}{dx_Fdp_T^2}
  =
  \int_0^\infty dp_T^2 \left(x_F^2+4\frac{m_T^2}{s}\right)^{-\frac{1}{2}}\frac{d^2\sigma}{dydp_T^2}
  \nonumber \\
  & = & 2\pi \int_0^\infty dp_T\,p_T \left(x_F^2+4\frac{m_T^2}{s}\right)^{-\frac{1}{2}} \left( E\frac{d^3\sigma}{dp^3}\right)\,,
  \nonumber
\end{eqnarray}
which, for $x_F=0$, simplifies to
\begin{equation}
  \left. \frac{d\sigma}{dx_F}\right|_{x_F=0}
  = \pi\sqrt{s}\int_0^\infty dp_T\,\frac{p_T}{m_T}
  \left.\left(E\frac{d^3\sigma}{dp^3} \right)\right|_{x_F=0}\,,
  \label{eq:integrated_xsec_xF0}
\end{equation}
where the quantity in parentheses on the r.h.s. of
Eq.~\eqref{eq:integrated_xsec_xF0} corresponds to Eq.~\eqref{eq:Lorentz_xsec}.
Furthermore, we shall repeat the evaluation of
Eq.~\eqref{eq:integrated_xsec_xF0} twice, respectively for the production of an
antiproton and of an antineutron in the final state. This task is non-trivial 
because, as we mentioned, the perturbative computation of 
the integrand in Eq.~\eqref{eq:integrated_xsec_xF0}, based on the collinear 
factorisation formula Eq.~\eqref{eq:Lorentz_xsec}, is valid only down to 
$p_T\sim 1$~GeV. A significant contribution to the integral, however, may 
originate from the region $0<p_T<1$~GeV.

In order to overcome this difficulty, we construct the integrand piecewise 
across the full $p_T$ range, separately for antiproton and antineutron 
production, combining a data-driven description at small $p_T$ with 
perturbative QCD predictions at large $p_T$, smoothly matched in an 
intermediate region. Specifically, we proceed as follows.

\begin{description}
\item[Antiproton.] 
  We consider three different regions of $p_T$. 

  \begin{enumerate}
  \item {\bf Region 1} ($p_T < 0.70$~GeV). This region lies well below
    the domain of validity of factorisation. In the absence of a theoretical 
    framework, we adopt a purely empirical treatment. We fit
    a two-parameter Gaussian model $f(p_T)=a\exp{(-b p_T^2)}$
    ($a,b\in \mathbb{R}$) to the NA49
    measurements~\cite{NA49:2009brx}. The choice of this functional form is
    the same as in~\cite{NA49:2009brx}. Although conceptually reminiscent of a
    transverse-momentum-dependent factorisation scheme, this parametrisation is
    purely phenomenological and the fitted parameters do not admit a direct 
    physical interpretation.

  \item {\bf Region 2} ($0.70\leq p_T\leq 1.30$~GeV). This intermediate 
    region provides a smooth interpolation between the small- and large-$p_T$ 
    behaviours of the cross section. We assume a functional form
    $f^\prime(p_T)=a^\prime\exp{(-b^\prime p_T^2 + c^\prime p_T)}$ 
    ($a^\prime,b^\prime,c^\prime \in\mathbb{R}$),
    with parameters fixed by requiring continuity and differentiability 
    at the boundaries with Regions~1 and~3.

  \item {\bf Region 3} ($p_T > 1.30$ GeV). In this region, we employ the 
  perturbative QCD predictions computed  as described in 
  Sect.~\ref{subsec:pert_comp}.

  \end{enumerate}

\item[Antineutron.] We consider two different regions of $p_T$.

  \begin{enumerate}
  \item {\bf Region 1} ($p_T < 1.30$~GeV). In contrast to the antiproton case, 
    no experimental measurements for the relevant cross section differential in 
    $p_T$ are available in this region. We therefore adopt a phenomenological 
    ansatz based on the observation that the cross section is dominated by 
    gluon fragmentation, which is largely insensitive to the hadron species.
    Under this assumption, the antineutron and antiproton cross sections are 
    expected to differ only mildly. We assume that this difference is constant
    in this region, and that it is equivalent to the difference between the 
    antiproton and antineutron cross sections computed at the boundary of the 
    region, $p_T=1.30$~GeV, which is of the order of 3\%. Our ansatz therefore 
    corresponds to assuming that the antineutron cross section is the same as
    the antiproton cross section shifted upwards by $~3$\%, and matched to the
    antiproton cross section computed in Region~2.

  \item {\bf Region 2} ($p_T > 1.30$~GeV). In this region we employ the 
    perturbative QCD predictions computed as detailed in
    Sect.~\ref{subsec:pert_comp}.
      
  \end{enumerate}

\end{description}

\begin{figure}[!t]
  \includegraphics[width=0.49\textwidth]{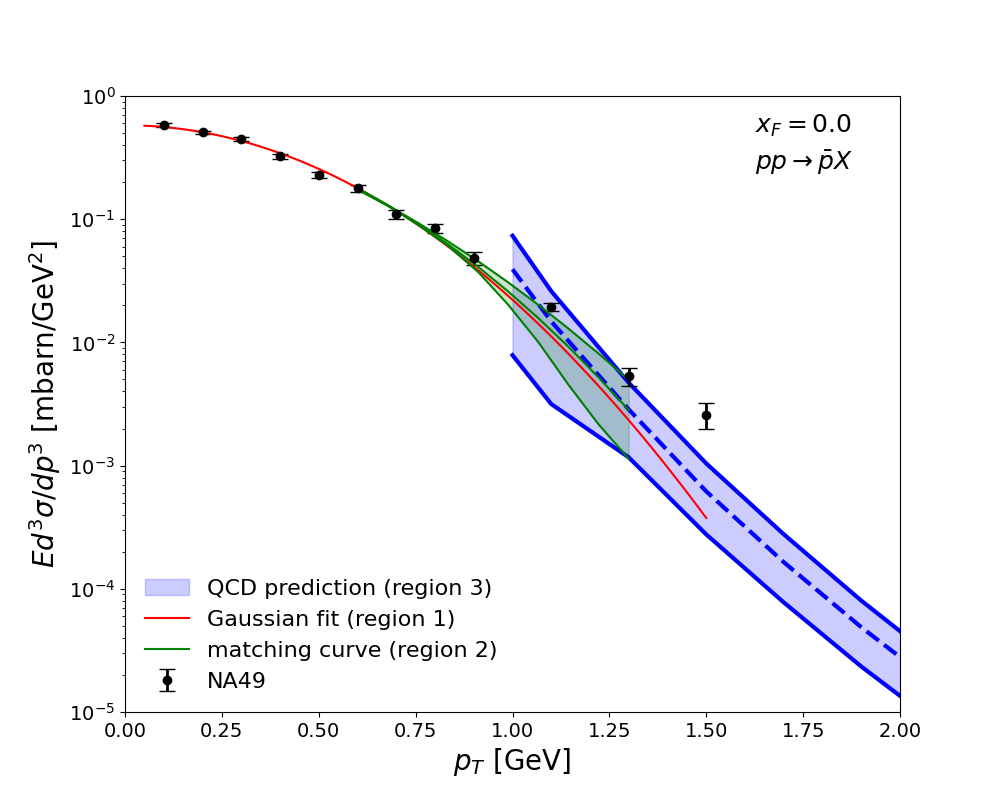}
  \includegraphics[width=0.49\textwidth]{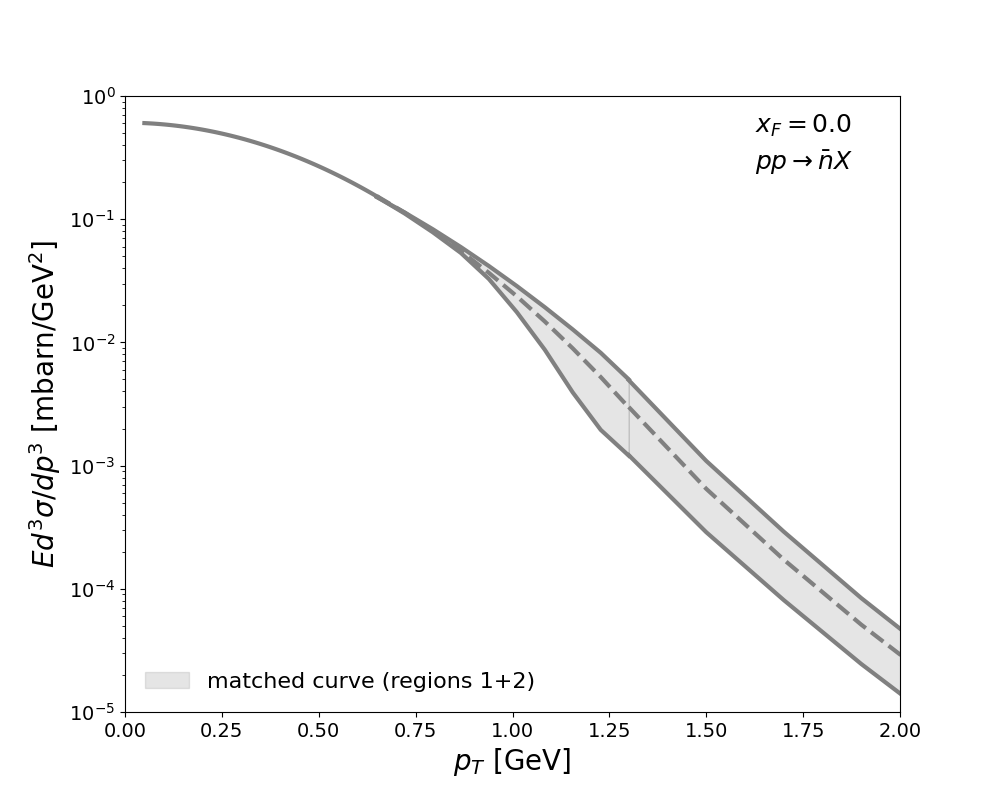}\\
  \caption{The Lorentz-invariant cross section, Eq.~\eqref{eq:Lorentz_xsec},
    for the inclusive production of antiprotons (left) and of antineutrons
    (right), reconstructed across the entire range of $p_T$ according to the
    procedure described in the text. For antiprotons, we also display the NA49
    measurements of~\cite{NA49:2009brx}.}
  \label{fig:matching}
\end{figure}

To determine the boundaries of the various regions, we consider the antiproton 
case and proceed as follows. For Region~3, we extend the perturbative 
computation down to the lowest possible value of $p_T$. As shown in 
Fig.~\ref{fig:antiprotons}, the perturbative prediction provides a satisfactory 
description of the experimental data down to $p_T \simeq 1.1$ GeV. We therefore 
choose this value as the matching point (see Fig.~\ref{fig:matching}). Having 
fixed this point, the width of the matching region (Region~2) is determined by 
requiring a smooth transition between the small- and large-$p_T$ behaviours. 
In practice, this is chosen such that it allows the interpolating function 
to acquire the appropriate curvature to match the two regimes smoothly, while 
remaining centred around $p_T \simeq 1$ GeV. The boundary between Region~1 and
Region~2 in the antineutron case is taken equal to the boundary between 
Region~2 and Region~3 in the antiproton case.

The result of our reconstruction procedure is displayed in
Fig.~\ref{fig:matching}, where we show the integrand cross section, smoothly
matched across the described regions of $p_T$, separately for antiproton and
antineutron production. The (tiny) uncertainty on the small-$p_T$ curve is
obtained by propagating the data uncertainty to the fitted parameters. The
larger uncertainty on the large-$p_T$ curve is the 68\% confidence level FF
uncertainty. The uncertainty on the intermediate-$p_T$ curve is obtained by
matching the two. The Gaussian fit to the small-$p_T$ antiproton NA49
measurements has a $\chi^2$ per degree of freedom of 1.42. A better fit,
with a lower value of the $\chi^2$, can be possibly obtained by using a
different parametrisation. However, we do not change it, and keep it the same
as in the NA49 analysis~\cite{NA49:2009brx}.

\subsection{Antineutron to antiproton integrated production yields}
\label{subsec:ratio_yields}

Using the reconstructed integrand discussed in Sect.~\ref{subsec:low_pT}, we
finally compute Eq.~\eqref{eq:integrated_xsec_xF0}. We also compute the ratio
between the antineutron to antiproton integrated cross sections, defined as
\begin{equation}  
  R =
  \left. \frac{d\sigma/dx_F (pp\to\bar n X)}{d\sigma/dx_F (pp\to \bar p X)}
  \right|_{x_F=0}\,.
  \label{eq:ratio}
\end{equation}
Our results are reported in Table~\ref{tab:integrated_xsec}, where we also
display the result reconstructed from~\cite{Fischer:2003xh} for the ratio $R$,
Eq.~\eqref{eq:ratio}. The (ff), (pdf), and (th) uncertainties come,
respectively, from the FF, PDF, and scale uncertainties of the perturbative
QCD computation, whereas the (model) uncertainty comes from the fitted
Gaussian model and from the matching curve. The total (tot) uncertainty is
the sum in quadrature of all of these components. Correlations are taken
into account in the ratio $R$, which lead to a large cancellation of PDF,
scale, and model uncertainties. 

\begin{table}[!t]
  \footnotesize
  \renewcommand{\arraystretch}{1.9}
  \begin{tabularx}{\textwidth}{lX}
  \toprule
  $d\sigma/dx_F|_{x_F=0}$ ($pp\to\bar p X$) [$\mu$barn]&  
  $0.462^{+0.029}_{-0.007}$ (pdf)
  $^{+0.256}_{-0.219}$ (ff)
  $^{+1.171}_{-0.342}$ (th)
  $^{+0.002}_{-0.053}$ (model)
  = $0.462^{+1.199}_{-0.331}$ (tot)\\
  $d\sigma/dx_F|_{x_F=0}$ ($pp\to\bar n X$) [$\mu$barn] &  
  $0.486^{+0.029}_{-0.007}$ (pdf)
  $^{+0.268}_{-0.231}$ (ff)
  $^{+1.227}_{-0.360}$ (th)
  $^{+0.001}_{-0.056}$ (model)
  = $0.486^{+1.256}_{-0.431}$ (tot)\\
  $R$ &  
  $1.048^{+0.000}_{-0.000}$ (pdf)
  $^{+0.002}_{-0.004}$ (ff)
  $^{+0.001}_{-0.010}$ (th)
  $^{+0.002}_{-0.001}$ (model)
  = $1.048^{+0.003}_{-0.011}$ (tot)\\
  $R$~\cite{Fischer:2003xh} &  
  $\sim 1.31$\\
  \bottomrule
  \end{tabularx}\\
  \vspace{0.3cm}
  \caption{The Lorentz-invariant cross section integrated over $p_T$ for
    $x_F=0$, Eq.~\eqref{eq:integrated_xsec_xF0}, computed for the inclusive
    production of antiprotons and antineutrons as explained in the text, and
    their ratio $R$, Eq.~\eqref{eq:ratio}. The corresponding value
    reconstructed from~\cite{Fischer:2003xh} is reported for comparison.
    The (ff), (pdf), and (th) uncertainties come, respectively, from the FF,
    PDF, and scale uncertainties of the perturbative QCD computation, 
    whereas the (model) uncertainty comes from the fitted Gaussian model and
    from the matching curve. The total (tot) uncertainty is the sum in
    quadrature of all of these components.}
  \label{tab:integrated_xsec}
\end{table}

From inspection of Table~\ref{tab:integrated_xsec}, the following remarks are
in order. First, we observe, consistently with Fig.~\ref{fig:antiprotons}, that
the expected value of the antiproton and antineutron integrated cross sections
are very close to each other. Second, again as expected from the error budget
displayed in Fig.~\ref{fig:antiprotons}, the uncertainty hierarchy is neat:
uncertainties coming from the large-$p_T$ QCD prediction dominate, in
particular scale uncertainties are the largest, ff uncertainties follow, and
pdf uncertainties are the smallest; uncertainties coming from the small-$p_T$
model, instead, are comparatively subdominant. We note that this fact is
relatively independent from the model used to fit the antiproton data or from
the ansatz used for the small-$p_T$ behaviour of the antineutron cross section.
The advantage of combining a small-$p_T$ data-driven model with large-$p_T$ QCD
predictions is therefore apparent: the former fixes the central value of the
integrated cross section; the latter provide a theory-grounded estimate of the
dominant sources of uncertainty. Similar considerations apply to the ratio $R$,
in which, despite large cancellations due to their correlations, non-negligible
uncertainties, in particular from scale variations, persist.
Our best result for the ratio $R$ is much smaller than that obtained 
in~\cite{Fischer:2003xh},
and not compatible with it within uncertainties. Instead, our result points
to a ratio close to unity, with a mild enhancement of antineutron over
antiproton production of, at most, a few percent. This is consistent with our 
observation that inclusive hadron production in \(pp\) collisions is dominated 
by gluon fragmentation, a partonic process that is largely insensitive to the
flavour difference between antiprotons and antineutrons.

\begin{figure}[!t]
  \centering
  \includegraphics[width=0.48\textwidth]{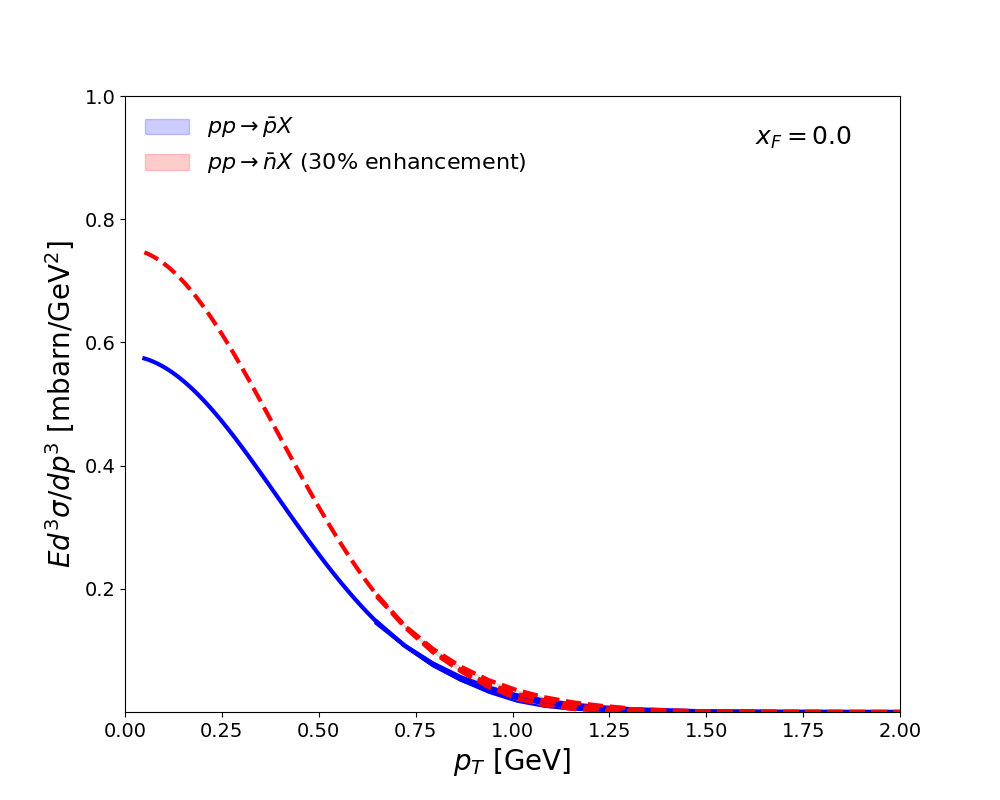}\\
  \caption{The reconstructed antiproton inclusive production cross section
    from Fig.~\ref{fig:matching} (solid) compared to the antineutron curve
    obtained by keeping the same large-$p_T$ tail, fixed by our QCD predictions,
    and a small-$p_T$ behaviour enhanced by 30\% (dashed).}
  \label{fig:rescaled_composite_new}
\end{figure}

We finally wonder whether our results can be reconciled with the apparently
inconsistent 30\% enhancement of antineutron to antiproton production cross
section reported by NA49. Uncertainties due to PDFs, FFs, or scale variations,
which we have estimated very conservatively, cannot account for this
discrepancy, at least in the region where QCD predictions are applicable.
To test whether a $30\%$ enhancement in antineutron over antiproton production
could arise from extreme, but still smooth, distortions of the cross section in 
the small-$p_T$ region, we artificially modify the antineutron small-$p_T$ 
proxy to force a $30\%$ enhancement of the integrated cross section.
This modification has no physical justification at this time, and it is solely 
intended as a diagnostic test. The perturbative
large-$p_T$ tail of the cross section is left unchanged, while the small-$p_T$
cross section is rescaled and then matched continuously to the large-$p_T$ 
prediction. The resulting artificially enhanced curve is compared to the 
antiproton reconstructed curve in Fig.~\ref{fig:rescaled_composite_new}.
To characterise the physical conditions under which such an enhancement could 
be realised, we note that achieving it would require a simultaneous and
rapid inversion of the relative contributions of gluon and quark fragmentation 
in a narrow range of transverse momentum. As $p_T$ decreases from $\sim 
1~\mathrm{GeV}$ to approximately zero, the quark fragmentation
component would need to become dominant, while the gluon contribution would
have to be strongly suppressed, in contrast with the behaviour observed at
larger $p_T$ (see Fig.~\ref{fig:channels}). In addition, a sizeable isospin 
asymmetry would need
to be introduced in this same kinematic region, enhancing $\bar{d}$ relative to
$\bar{u}$ production in $pp$ collisions well beyond standard expectations.
Only if both of these conditions are realised can the $\bar{n}/\bar{p}$ ratio
significantly exceed unity, without modifying the large-$p_T$ cross section 
controlled by perturbative QCD. Within the present framework, there is no 
independent evidence supporting such a scenario. Our analysis therefore
suggests that the discrepancy with NA49 may originate either from genuinely
non-perturbative dynamics beyond the reach of collinear factorisation, or from
experimental systematics not fully captured in the published result.

We end this section by noting that differences between cross sections for other
hadrons remain of the same order as the ones observed here between antiprotons
and antineutrons. For instance, NA49 measured the Lorentz-invariant cross
section as a function of $p_T$ in different bins of $x_F$ for the inclusive
production of positively and negatively charged pions~\cite{NA49:2005qor}.
They found no evidence of differences between positive and negative pion
production cross sections at $x_F=0$ within their experimental uncertainties,
whereas they found larger positive pion yields in comparison to
negative pions for larger values of $x_F$. Differences increase at small $p_T$,
a fact in support of our hypothesis according to which some non-perturbative
mechanism occurring in that kinematic region enhances the difference between
the inclusive production of different hadron yields. We review these results,
and compare them to our QCD theoretical predictions, in
Appendix~\ref{app:pions}. We find a generally good agreement between
experimental data and theory predictions, where applicable, within their rather
large uncertainties. We also observe an increase of the difference between
positive and negative pion production cross sections as $x_F$ increases,
although the statistical significance of these differences remains mild due to
large FF uncertainties. In this respect, the NA49 data may be used as a
constraint on FFs themselves, once the large scale uncertainties are reduced
thanks to the inclusion of NNLO perturbative corrections. We leave this further
line of research to future work.

\section{Conclusions}
\label{sec:conclusions}

In this paper we have studied inclusive antiproton and antineutron production
in \(pp\) collisions within a framework based on perturbative QCD matched
to a phenomenological description of the small-$p_T$ region. This approach
combines the precision of available small-$p_T$ measurements with a
theory-anchored description of the large-$p_T$ behaviour that allows us to 
systematically assess model and theoretical uncertainties. Our goal was to
reassess the hadronic input relevant to CR antiproton production, and
in particular to revisit the longstanding claim by the NA49 experiment of a
$\sim 30\%$ excess of antineutron over antiproton production.

We first considered the Lorentz-invariant cross section for inclusive
antiproton production and discussed its computation in collinear factorisation
in terms of perturbative hard coefficients, proton PDFs, and hadron FFs. 
We showed how measurements performed by collider experiments such as NA49 
and ALICE probe the kinematic region relevant to CR antiprotons. We analysed the
partonic origin of the cross section and found that, in the kinematic region of
interest, it is largely dominated by gluon fragmentation. This observation 
suggests that inclusive antiproton and antineutron production should be rather 
similar, since the two channels differ only through the quark fragmentation 
component.

We then computed perturbative predictions for inclusive antiproton production
at the NA49 centre-of-mass energy and compared them to the available data. In
the region where perturbative QCD is applicable, namely for
$p_T \gtrsim 1~\mathrm{GeV}$, the predicted cross section provides a reasonable
description of the measurements within uncertainties. We also separately
assessed the PDF, FF, and scale uncertainties, finding the expected hierarchy:
PDF uncertainties are small, FF uncertainties are larger, and scale
uncertainties dominate the error budget at such relatively small $p_T$.

Since the NA49 claim concerns the cross section integrated over the full range
of $p_T$, we complemented the perturbative calculation with a phenomenological
reconstruction of the small-$p_T$ region. For antiprotons, this reconstruction
was based on a Gaussian fit to the NA49 data, smoothly matched to the
perturbative prediction through an intermediate region. For antineutrons, where
no differential measurements in $p_T$ are available, we constructed a proxy for
the small-$p_T$ behaviour by assuming that the difference with antiprotons
remains controlled by the same mechanism observed at large $p_T$, namely the
subleading quark-fragmentation component. This yields a smooth continuation of
the antiproton reconstruction shifted by only a few percent. We found that
the theory uncertainties related to the large-$p_T$ QCD predictions dominate
the uncertainty budget of the cross section, with model uncertainties
related to the small-$p_T$ parametrisation or proxy being comparatively
negligible. Most importantly, large-$p_T$ uncertainties are independent from the
choices made to model the small-$p_T$ behaviour of the cross section.

Using this hybrid construction, we evaluated the $p_T$-integrated cross
sections at $x_F=0$ for both inclusive antiproton and inclusive antineutron
production, and their ratio. Our best result is
\[
R = \frac{d\sigma/dx_F(pp\to \bar{n}X)}{d\sigma/dx_F(pp\to \bar{p}X)}
\bigg|_{x_F=0}
= 1.048^{+0.003}_{-0.011},
\]
where the effect of correlations has been fully taken into account, leading to 
significant uncertainty cancellations. Despite these cancellations, residual 
uncertainties, in particular from FFs and scale variations, remain 
non-negligible. Our result is therefore much smaller than the $\sim 30\%$ 
enhancement quoted by NA49, and instead points to a ratio close to unity, 
with at most a mild enhancement of antineutron over antiproton production at 
the level of a few percent.

We also investigated whether the NA49 result could be reconciled with our 
framework. To this end, we artificially modified the antineutron
cross section in the small-$p_T$ region while keeping the perturbative large-
$p_T$ tail unchanged, and determined the distortion required to reproduce a 
$30\%$ enhancement after integration over $p_T$. 
We found that such a construction is mathematically possible, but only at the 
cost of introducing sizeable differences between antiproton and antineutron 
production precisely in the small-$p_T$ region. Within the framework adopted in 
this work, there is no independent evidence for such a  large effect.
Our analysis therefore suggests that the discrepancy with NA49 may originate
either from genuinely non-perturbative dynamics beyond the reach of collinear
factorisation, or from experimental systematics not fully captured in the
published result.

Finally, the comparison presented in Appendix~\ref{app:pions}
supports the broader picture emerging from our study. Inclusive charged-pion 
production at NA49 is also reasonably well described by perturbative QCD in the
region where the calculation is expected to apply, while larger differences
between hadron species appear as one moves towards smaller transverse momenta.
This observation is consistent with the conclusion that the dominant unresolved 
hadronic uncertainty in the present problem is concentrated in the
non-perturbative small-$p_T$ domain.

The theoretical framework developed here can be refined as new data become 
available. To this end, the AMBER experiment at CERN collected its first data
in 2023--2024 specifically targeting the kinematic regime analysed in this
work~\cite{Adams:2018pwt,Giordano:2025ooi,Giordano:2025nwb}. AMBER measurements
span centre-of-mass energies of $\sqrt{s}=10.7$--$21.7~\mathrm{GeV}$, 
including $pp$ and \(pD\) collisions at several beam energies, with projected 
$\sim 5\%$ uncertainties on cross sections and $\sim 10\%$ constraints on 
isospin asymmetries. In parallel, the LHCb experiment is exploiting the SMOG
fixed-target configuration to study hadronic interactions with light nuclear
targets~\cite{BoenteGarcia:2024kba}. Measurements with deuterium targets will
constrain differences between $pp$ and $pn$ collisions, providing direct input
on the relation between antiproton and antineutron production. Together, these
measurements will provide a stringent experimental test of the framework
presented here in the energy range most relevant for the interpretation of
precise CR antiproton data.

\section*{Acknowledgements}
\label{sec:acknowledgements}
M.D.M. and F.D.~acknowledge support from the research grant
{\sl TAsP (Theoretical Astroparticle Physics)} funded by Istituto Nazionale di
Fisica Nucleare (INFN). M.B. ad A.S. acknowledge partial support by the
European Union "Next Generation EU" program through the Italian PRIN 2022 grant
n.~20225ZHA7W. E.R.N. was supported by the Italian Ministry of University and
Research (MUR) through the "Rita Levi-Montalcini" Program. J.R.W. acknowledges
support from INFN Turin, Grant No. 25864.

\appendix
\section{Validation against LHC ALICE data}
\label{app:ALICE}

In this Appendix, we validate our perturbative QCD computation of the
Lorentz-invariant cross section, Eq.~\eqref{eq:Lorentz_xsec}, against the
antiproton ALICE measurements~\cite{ALICE:2020jsh} included in the {\sc NPC23}
analysis~\cite{Gao:2024dbv}. The measurement corresponds to the ratio 
between Lorentz-invariant cross sections for the production of the sum of
protons and antiprotons at centre-of-mass energies of 13 and 7~TeV.
The cross sections are integrated over the rapidity $y$ in the range $|y|<0.5$.

Using consistently their PDFs ({\sc CT24nlo}~\cite{Dulat:2015mca}) and FFs
({\sc NPC23}~\cite{Gao:2024dbv}), our predictions, which we otherwise compute
as detailed in Sect.~\ref{subsec:pert_comp}, actually become postdictions, that
must match the data as in~\cite{Gao:2024dbv}. We also verify that our default
non-perturbative input parton sets, {\sc NNPDF4.0}~\cite{NNPDF:2021njg} for
PDFs and  {\sc NNFF1.0}~\cite{Bertone:2017tyb} for FFs, despite not including
the aforementioned measurements, provide a good description of them. This
sanity check corroborates the reliability of our chosen non-perturbative input
to correctly describe the cross section over a broad range of $p_T$ values.

\begin{figure}[!t]
  \centering
  \includegraphics[width=0.49\textwidth]{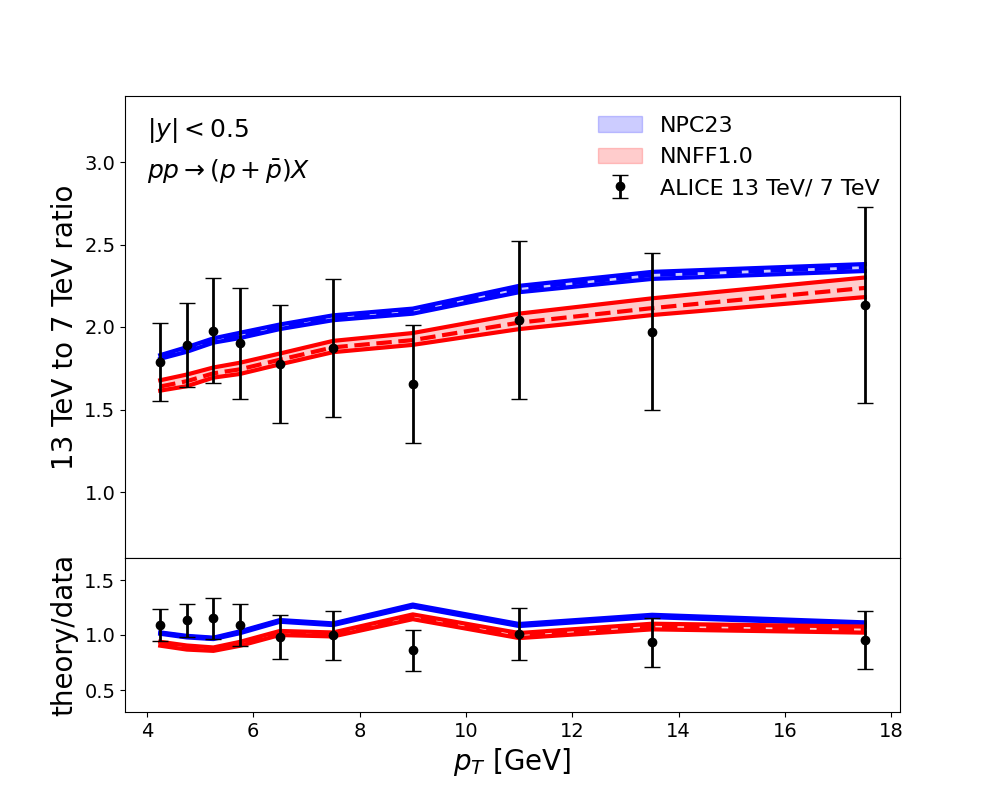}\\  
  \caption{Comparison between the antiproton ALICE
    measurements~\cite{ALICE:2020jsh} included in the {\sc NPC23}
    analysis~\cite{Gao:2024dbv}, and the theoretical postdictions, computed as
    detailed in Sect.~\ref{subsec:pert_comp}, using consistently their PDFs
    and FFs, {\sc CT24nlo}~\cite{Dulat:2015mca} and
    {\sc NPC23}~\cite{Gao:2024dbv}. Theoretical predictions, obtained with our
    default non-perturbative input parton sets,
    {\sc NNPDF4.0}~\cite{NNPDF:2021njg} for PDFs and
    {\sc NNFF1.0}~\cite{Bertone:2017tyb} for FFs are also shown.
    The measurements correspond to the ratio of the inclusive production of the
    sum of protons and antiprotons at a centre-of-mass energy of 13~TeV to the
    sum of protons and antiprotons at a centre-of-mass energy of
    7~TeV~\cite{ALICE:2020jsh}. Theoretical uncertainties
    correspond to 68\% confidence level and account for FF uncertainties only
    (see text for details). The uncertainties on the experimental data are the
    sum in quadrature of statistical and systematic uncertainties. The lower
    panel shows the results normalised to the experimental data.}
  \label{fig:validation}
\end{figure}

Figure~\ref{fig:validation} displays a comparison between the experimental data
and the theoretical postdictions (for {\sc NPC23}) or predictions
(for {\sc NNFF1.0}). The comparison is shown for values of the
transverse momentum of the produced hadrons, $p_T$, larger than 4~GeV,
consistently with the cut applied in the {\sc NPC23}
analysis~\cite{Gao:2024dbv}. The uncertainties on the {\sc NPC23} theoretical
postdictions are Hessian uncertainties accounting for the FF uncertainty only,
which we determine according to Eq.~(E1) in~\cite{Gao:2024dbv}, whereas the
uncertainties on the {\sc NNFF1.0} theoretical predictions are Monte Carlo
uncertainties, accounting for FF uncertainties only also in this case. Bands
correspond to 68\% confidence levels in both cases. The uncertainties on the
experimental data are the sum in quadrature of statistical and systematic
uncertainties. As in~\cite{Gao:2024dbv}, possible additional normalisation
uncertainties are not shown. 

A comparison between Fig.~\ref{fig:validation} and Fig.~17 in~\cite{Gao:2024dbv}
reveals the same pattern of data-theory comparison. We have checked that
similar results hold for other ALICE measurements included in the {\sc NPC23}
analysis, namely the ratio of the inclusive production of the sum of protons
and antiprotons to the sum of positively and negatively charged pions at
centre-of-mass energies of 2.76~TeV~\cite{ALICE:2014juv},
5.02~TeV~\cite{ALICE:2019hno}, and 13~TeV~\cite{ALICE:2020jsh}. In particular,
we recover the same shape, including notably slopes in data normalised to
theory postdictions for ratios of protons to pions at all three centre-of-mass
energies, and a trend of normalisation shift, in particular at large $p_T$,
at 2.76 and 13~TeV. We interpret the fact that our numerical framework
reproduces the features of the {\sc NPC23} analysis as a validation of its
reliability. The theoretical predictions obtained with our default
non-perturbative input describes the data as nicely as {\sc NPC23}, albeit
with slightly larger uncertainties. This sanity check demonstrates that FFs
are under good control over a wide range of $p_T$ values.

\section{Charged pion production at NA49}
\label{app:pions}

\begin{figure}[!t]
  \centering
  \includegraphics[width=0.49\textwidth]{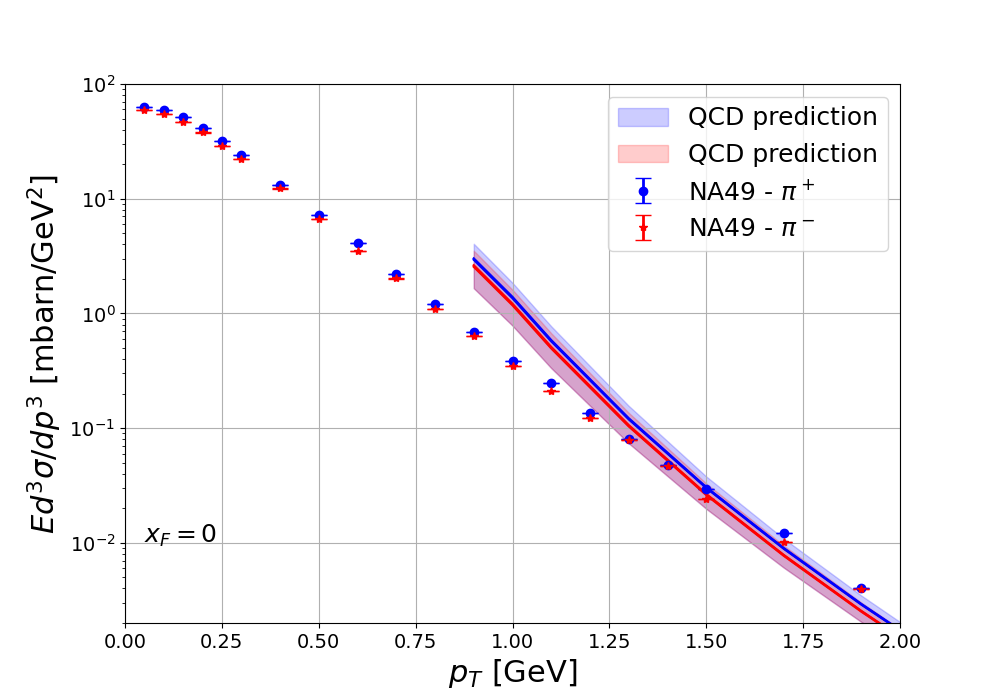}
  \includegraphics[width=0.49\textwidth]{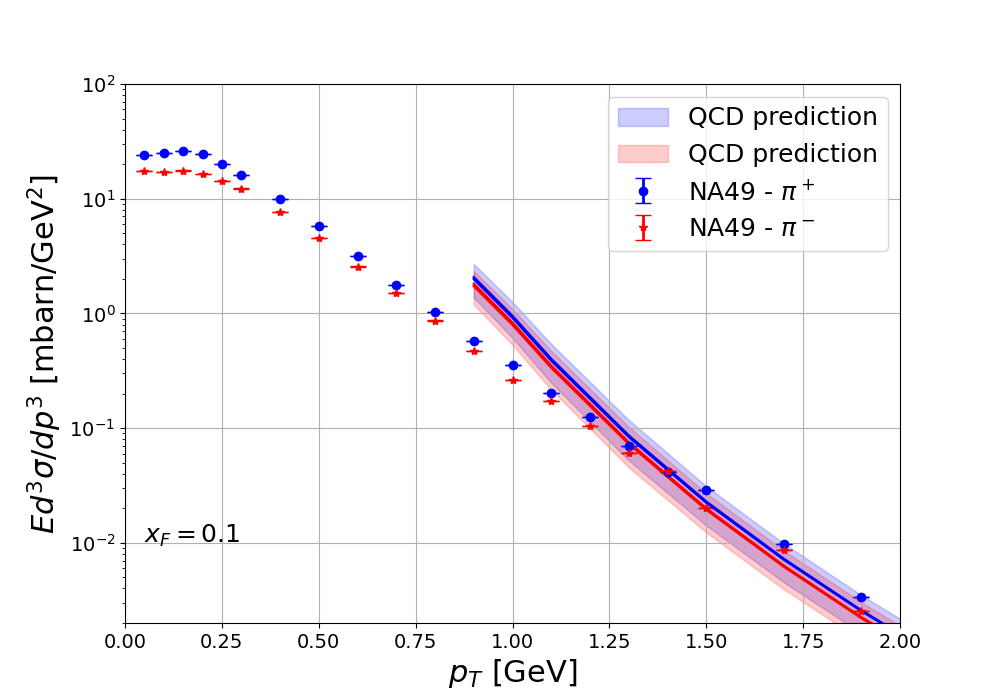}\\
  \includegraphics[width=0.49\textwidth]{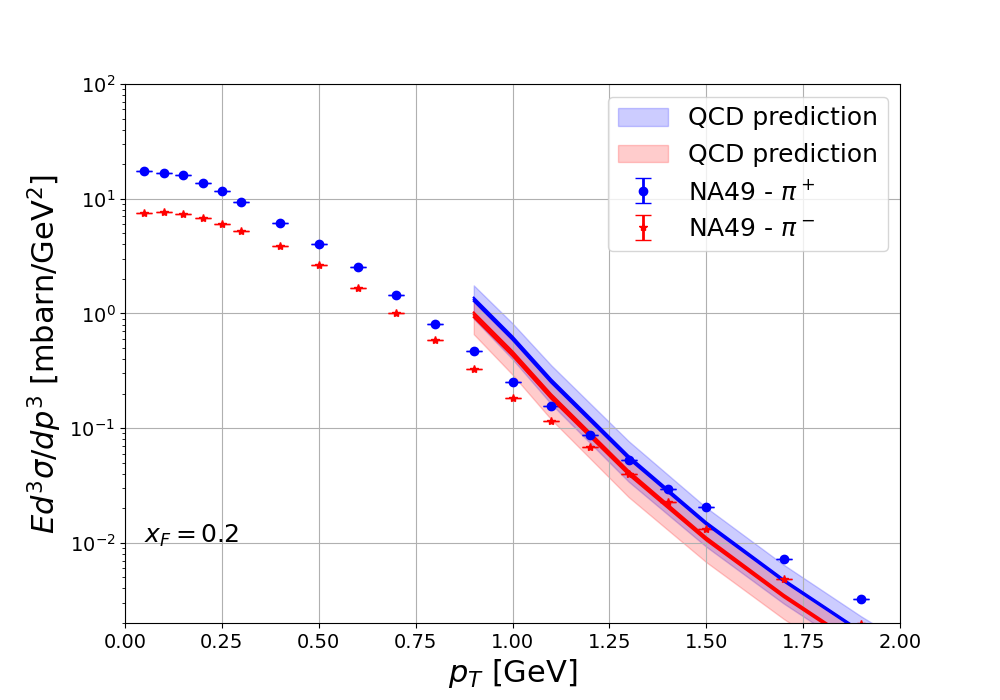}
  \includegraphics[width=0.49\textwidth]{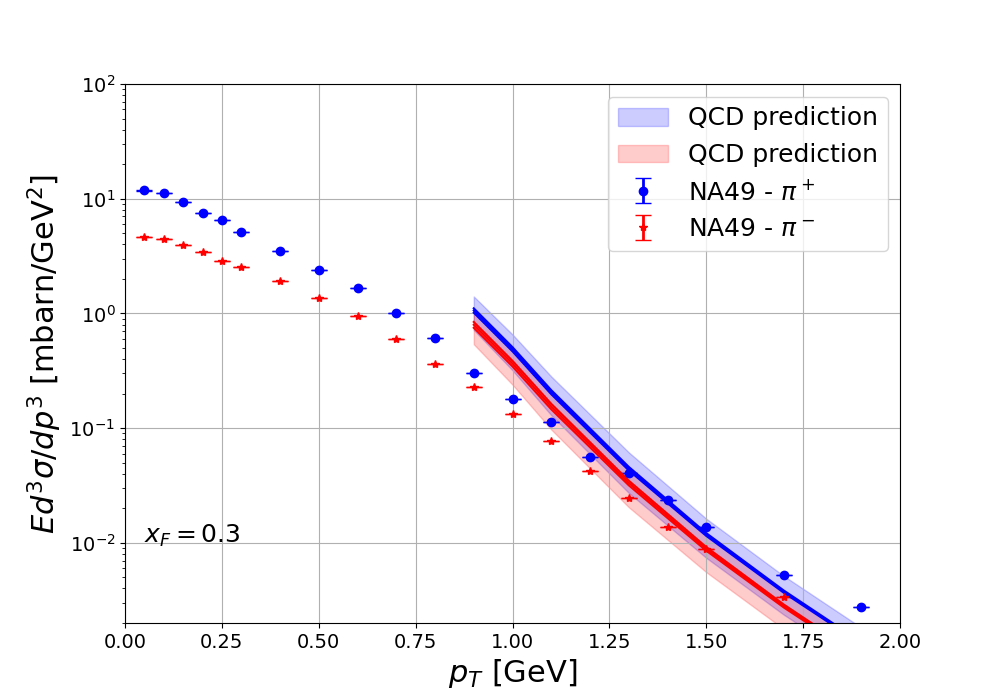}\\
  \caption{Comparison of theoretical predictions, computed as described in
    Sect.~\ref{subsec:pert_comp}, for the inclusive production cross section
    of negatively and positively charged pions in \(p p \) collisions to
    the corresponding measurements reported by NA49 in~\cite{NA49:2005qor}.
    Measurements correspond to a centre-of-mass energy of 17.2~GeV and are
    differential in the transverse momentum of the detected pion, $p_T$, for
    fixed values of $x_F$. We display results for $x_F=0.0, 0.1, 0.2, 0.3$
    and truncate theoretical predictions in the range of applicability of
    perturbative QCD ($p_T>1$~GeV). The uncertainties on theoretical
    predictions correspond to the 68\% confidence level FF uncertainty only.
    The uncertainties on the data are the sum in quadrature of statistical and
    systematic uncertainties.}
  \label{fig:pions}
\end{figure}

In this Appendix, we compute the Lorentz invariant cross section,
Eq.~\eqref{eq:Lorentz_xsec}, for the inclusive production of positively and
negatively charged pions in \(p p \) collisions. The details of the
computation are as in Sect.~\ref{subsec:pert_comp}, but we now use the NNFF1.0
$\pi^+$ and $\pi^-$ FF sets. We compare our theoretical predictions, in the
range of applicability of perturbative QCD ($p_T>1$~GeV), to the NA49
measurements presented in~\cite{NA49:2005qor}. These correspond to a
centre-of-mass energy of 17.2~GeV and are differential in the
transverse momentum of the detected pion, $p_T$, for fixed values of $x_F$.

In Fig.~\ref{fig:pions} we display the comparison for a subset of the measured
values of $x_F$, namely 0.0, 0.1, 0.2, and 0.3. The uncertainties on theoretical
predictions correspond to the 68\% confidence level FF uncertainty only. We
have checked that PDF and scale uncertainties are similar to those displayed in
Fig.~\ref{fig:antiprotons} for antiprotons. The uncertainties on the data are
the sum in quadrature of statistical and systematic uncertainties.

Our theoretical predictions, where applicable, provide a generally good
description of the experimental measurements, albeit with uncertainties that
are much larger than the data uncertainties. We observe that there is no
evidence for a difference between cross sections for positive and negative
pions at $x_F=0$ within experimental uncertainties. Larger differences appear
as $x_F$ increases, with positive pion yields being larger than negative pion
yields, in particular at low values of $p_T$. This fact supports our hypothesis
according to which some non-perturbative mechanism occurring in that kinematic
region enhances the difference between the inclusive production of different
hadron yields. At large $p_T$, this trend is less visible in the data, although
it is partly recovered by our theoretical predictions. However, the statistical
significance of these differences remains mild due to large uncertainties.
In this respect, the NA49 data may be used as a constraint on FFs themselves,
once the large scale uncertainties are reduced thanks to the inclusion of NNLO
perturbative corrections. We leave this further line of research to future work.

\bibliographystyle{utphys}
\bibliography{antiproton}

\end{document}